\documentclass[iop]{emulateapj}

\newcommand{\myemail}{mawatari@las.osaka-sandai.ac.jp}

\slugcomment{Not to appear in Nonlearned J., 45.}

\shorttitle{Galaxy-DLA at $z = 3.3$}
\shortauthors{Mawatari et al.}

\begin{document}


\title{Discovery of a Damped Ly$\alpha$ Absorber at $z = 3.3$\\ along a galaxy sight-line in the SSA22 field}


\author{K. Mawatari\altaffilmark{1}, A. K. Inoue\altaffilmark{1}, K. Kousai\altaffilmark{2}, T. Hayashino\altaffilmark{2}, R. Cooke\altaffilmark{3}, J. X. Prochaska\altaffilmark{3}, T. Yamada\altaffilmark{4}, and Y. Matsuda\altaffilmark{5}}
\email{\myemail}


\altaffiltext{1}{College of General Education, Osaka Sangyo University, 3-1-1, Nakagaito, Daito, Osaka, 574-8530, Japan}
\altaffiltext{2}{Research Center for Neutrino Science, General School of Science, Tohoku University, Aoba, Aramaki, Aoba-ku, Sendai, Miyagi, 980-8578, Japan}
\altaffiltext{3}{Department of Astronomy and Astrophysics, UCO/Lick Observatory, University of California, 1156 High Street, Santa Cruz, CA 95064, USA}
\altaffiltext{4}{Astronomical Institute, Tohoku University, Aoba, Aramaki, Aoba-ku, Sendai, Miyagi, 980-8578, Japan}
\altaffiltext{5}{National Astronomical Observatory of Japan, Osawa 2-21-1, Mitaka, Tokyo 181-8588, Japan}


\begin{abstract}
Using galaxies as background light sources to map the Ly$\alpha$ absorption lines is a novel approach to study Damped Ly$\alpha$ Absorbers (DLAs). We report the discovery of an intervening $z = 3.335 \pm 0.007$ DLA along a galaxy sight-line identified among 80 Lyman Break Galaxy (LBG) spectra obtained with our Very Large Telescope/Visible Multi-Object Spectrograph survey in the SSA22 field. The measured DLA neutral hydrogen (H~{\sc i}) column density is $\log (N_{\rm HI}/{\rm cm^{-2}}) = 21.68 \pm 0.17$. The DLA covering fraction over the extended background LBG is $> 70$\,\% ($2 \sigma$), yielding a conservative constraint on the DLA area of $\ga 1$\,kpc$^2$. Our search for a counterpart galaxy hosting this DLA concludes that there is no counterpart galaxy with star formation rate (SFR) larger than a few\,M$_{\odot}$\,yr$^{-1}$, ruling out an unobscured violent star formation in the DLA gas cloud. We also rule out the possibility that the host galaxy of the DLA is a passive galaxy with $M_* \gtrsim 5 \times 10^{10}$\,$M_{\sun}$ or a heavily dust-obscured galaxy with $E(B-V) \gtrsim 2$. The DLA may coincide with a large-scale overdensity of the spectroscopic LBGs. The occurrence rate of the DLA is compatible with that of DLAs found in QSO sight-lines.
\end{abstract}





\section{Introduction}

Damped Ly$\alpha$ Absorbers (DLAs) are neutral hydrogen (H~{\sc i}) gas clouds with a high column density ($N_{\rm HI} > 2 \times 10^{20}$\,cm$^{-2}$; \citealt{Wolfe+86}) typically identified in the spectra of bright background objects. DLAs at high redshift ($z \sim 3$) contain a significant fraction of H~{\sc i} gas in the universe and their gas mass is $\sim 20 - 50$\,\% of the present-day stellar mass \citep{Lanzetta+95,Storrie-Lombardi+00}. Therefore, investigating the nature of DLAs and their link with stellar components is clearly important to understand the baryon physics on galaxy formation. 

% QSO-DLA
Traditionally QSOs have been used as background light sources to study DLAs (we call DLAs in QSO sight-lines as ``QSO-DLAs'' throughout this paper). Their extremely bright flux allows DLAs to be identified over a broad redshift range up to $z \sim 5$, even with wide and shallow surveys \citep{Prochaska+05,Prochaska+09,Rafelski+12,Crighton+15}. On the other hand, the limited information that one gathers along the quasar line of sight does not reveal the size and structure of the H~{\sc i} gas, which obscures the true nature of DLAs. Both the rotational motion of disk galaxies and the combination of infall and random motion of pre-galactic clumps can explain the observed kinematic properties of DLAs \citep{Haehnelt+98,Prochaska+98}.

Direct identification of DLA host galaxies in emission is a straightforward way to investigate the link between H~{\sc i} gas and stellar components in DLAs. Imaging surveys of DLA counterpart galaxies in the local universe have revealed a wide variety of galaxies hosting DLAs \citep{Chen+03,Rao+03}. At $z > 2$, a small number of ($= 10 - 20$) galaxies associated with DLAs has been found so far \citep{Krogager+12,Peroux+12}. A small impact parameter of counterpart galaxies from the background QSOs ($b \lesssim 25$\,kpc; \citealt{Krogager+12}) and high contrast between their brightness make it difficult to detect faint continuum emission from the counterpart galaxies, although some authors overcame these difficulties by searching emission lines (e.g., \citealt{Fynbo+10,Peroux+11,Peroux+12,Noterdaeme+12a}) or by using a sophisticated method (double-DLA technique; \citealt{OMeara+06,Christensen+09,Fumagalli+10,Fumagalli+14,Fumagalli+15}). 

Sometimes DLAs are also identified in spectra of gamma-ray bursts (GRB-DLAs; \citealt{Vreeswijk+04}). GRB-DLAs have a significant merit in searching for galaxies hosting intervening DLAs because GRB afterglows become fainter and the contrast with counterpart galaxies increases with time.

%gal-DLA; new population
In this paper, we report a new type of DLAs, ``gal-DLAs,'' which is identified in the spectra of normal galaxies. Using galaxies as a background sources generally benefits us in the following ways: (i) we can search for the counterpart galaxies at a smaller impact parameter in any wavelength because of low brightness contrast between the background and counterpart galaxies, and (ii) extended background sources enable us to resolve the DLA absorption features spatially or to investigate DLA covering factors over the background sources by measuring the residual flux in the Ly$\alpha$ trough. Also, another gal-DLA has just been reported by \citet{Cooke+15}, and it is expected that a large number of gal-DLAs will be identified in archival and future large spectroscopic survey data. The gal-DLAs will become a key population to investigate the neutral gas reservoirs at high redshift. 

We briefly describe the observations in Section~2, and discuss the properties of the identified gal-DLA in Section~3. We use the AB magnitude system \citep{OkeGunn83} and adopt a cosmology with $H_{0}=70.4$ km s$^{-1}$ Mpc$^{-1}$, $\Omega_{M}=0.272$, and $\Omega_{\Lambda}=0.728$ \citep{komatsu11}. We also adopted a \citet{Chabrier+03} initial mass function (IMF) with the mass range of $0.1\,M_{\sun} - 100\,M_{\sun}$ to estimate the star formation rate (SFR) and the stellar mass.

\section{Observation And Data}

%% In a manner similar to \objectname authors can provide links to dataset
%% hosted at participating data centers via the \dataset{} command.  The
%% second curly bracket argument is printed in the text while the first
%% parentheses argument serves as the valid data set identifier.  Large
%% lists of data set are best provided in a table (see Table 3 for an example).
%% Valid data set identifiers should be obtained from the data center that
%% is currently hosting the data.
%%
%% Note that AASTeX interprets everything between the curly braces in the 
%% macro as regular text, so any special characters, e.g. "#" or "_," must be 
%% preceded by a backslash. Otherwise, you will get a LaTeX error when you 
%% compile your manuscript.  Special characters do not 
%% need to be escaped in the optional, square-bracket argument.

We performed spectroscopic observations of photometrically selected Lyman Break Galaxies (LBGs) with the Visible Multi-Object Spectrograph (VIMOS; \citealt{LeFevre+03}) on the Very Large Telescope (VLT). 

The target objects for spectroscopy were selected in the SSA22 field, using the $u^*$ band image \citep{Kousai+11} taken with CFHT/Megacam \citep{Boulade+03} and the $V$, $R_c$, and $i'$ band images \citep{Hayashino+04} taken with Subaru/Suprime-Cam \citep{Miyazaki+02}. We applied the following LBG selection to the objects detected in the $R_c$ band image with the $2$\,arcsec diameter aperture photometry: (i) $23.9 \leqslant R_c \leqslant 25.4$, (ii) $(u^*-V)-1.8(V-R_c) \geqslant 1.1$, and (iii) $R_c - i' \leqslant 0.3$. Star-forming galaxies at $z \ga 3$ are expected to be selected with these criteria \citep{Kousai+11}. 

\begin{figure}[]
%\begin{figure}[t]
\begin{center}
\includegraphics[width=1.0\linewidth, angle=0]{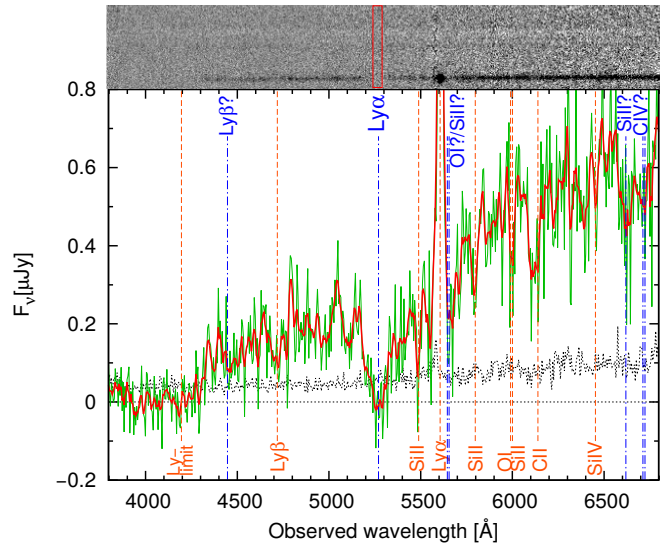}
\caption{Two-dimensional (top) and one-dimensional (bottom) spectra of the LBG at $z = 3.604 \pm 0.008$ in which the DLA at $z = 3.335 \pm 0.007$ can be seen. The red box superposed on the top panel shows the region where we searched for the Ly$\alpha$ emission line from the counterpart galaxy. In the bottom panel, the thick red and thin green lines are the spectrum with and without a 5 pixel box-car smoothing, respectively. The dotted black line is the error spectrum expected from the root-mean-square spectrum of the sky-subtracted background. The vertical dotted-dashed blue lines show the significant Ly$\alpha$ absorption and possible metal absorption lines by the gal-DLA, while the dashed orange lines correspond to the emission/absorption features of the background LBG itself. \label{fig:DLAspec}}
\end{center}
\end{figure}

Our VIMOS observations in 2008 (VIMOS08; the program ID of 081.A-0081(A), PI: A. K. Inoue) is comprised of 163 LBGs satisfying the above criteria in two fields of view (FOVs): one is centered at $(\alpha,\delta) = (22^h17^m31.9^s,+00{\arcdeg}24{\arcmin}29.7{\arcsec})$ and the other is centered at $(\alpha,\delta) = (22^h17^m39.1^s,+00{\arcdeg}11{\arcmin}00.7{\arcsec})$. Total on-source integration time is 14,080 s for each FOV. The data were acquired with a spectral resolution of $R \simeq 180$ and a pixel scale of 5.3\,\AA/pix. 

We reduced the raw data with the VIMOS pipeline\footnote{http://www.eso.org/sci/software/pipelines/vimos/} and NOAO IRAF\footnote{http://iraf.noao.edu} (see T. Hayashino et al. 2016, in preparation and \citealt{Kousai+11} for details). From the reduced two-dimensional spectral images, we extracted 4\,pixels ($= 0.82$\,arcsec) in the spatial direction to trace the object continuum and summed them to produce the one-dimensional spectra. These are further smoothed with a 5 pixel box-car kernel to increase the signal-to-noise ratio (S/N) for each spectrum, where 5 spectral pixels nearly correspond to the VIMOS resolution. We searched for spectral features such as the Ly$\alpha$ emission/absorption line and metal absorption lines by eye in the smoothed spectra, from which we estimated the systemic redshifts following the calibration formulae of \citet{Adelberger+05}. In this study we focus on the 80 LBGs with reliable redshifts (classes Ae, Aa, and B in T. Hayashino et al. 2016, in preparation). More details about the observations and reduction are described in \citet{Kousai+11}, \citet{Inoue+11}, and T. Hayashino et al. (2016, in preparation).

Among the 80 LBG spectra, we serendipitously discovered a strong, intervening Ly$\alpha$ absorption feature in a $z = 3.604 \pm 0.008$ LBG at $(\alpha,\delta) = (22^h17^m06.9^s,+00{\arcdeg}05{\arcmin}39.0{\arcsec})$. The spectrum of this galaxy is shown in Figure~\ref{fig:DLAspec}. While our visual identification of this DLA does not come from a systematic survey of DLAs, we identify no other DLA candidate as strong as the example in Figure~\ref{fig:DLAspec}.

%% In this section, we use  the \subsection command to set off
%% a subsection.  \footnote is used to insert a footnote to the text.


%% This section also includes several of the displayed math environments
%% mentioned in the Author Guide.

\section{Results And Discussion}

\subsection{The gal-DLA gas content} \label{sec:DLAgas}

\begin{figure}[]
%\begin{figure}[t]
\begin{center}
\includegraphics[width=0.9\linewidth, angle=90]{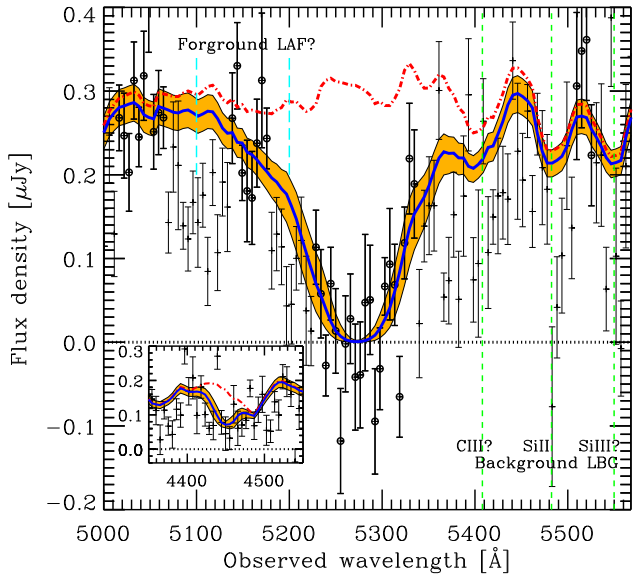}
\caption{An example of the Voigt function fit for the gal-DLA spectrum. The flux points used for the Voigt function fit are shown by the thick black bars with circles, which are selected avoiding the possible absorption lines from the background LBG (vertical green dashed line) and foreground H~{\sc i} absorbers (vertical cyan dashed line). The best-fit spectrum and the acceptable fit within the $1 \sigma$ confidence level is shown by the blue bold line and orange shading, respectively. The best-fit continuum spectrum is shown by the red dotted-dashed line. This example yields the best-fit H~{\sc i} column density as $\log(N_{\rm HI}/cm^{-2}) = 21.66$. The small panel embedded in the bottom left is the same but around the possible Ly$\beta$ absorption from the gal-DLA, where no flux point is used for the fit. \label{fig:voigtfit}}
\end{center}
\end{figure}

First, we fit the DLA absorption at $\lambda = 5000 - 5550$\,\AA\ with the Voigt function to estimate the redshift and H~{\sc i} column density. We used the composite spectrum, which are made by stacking the 39 LBG spectra with the Ly$\alpha$ emission line (class Ae in T. Hayashino et al. 2016, in preparation), as a continuum. We carefully selected the flux points used for the fit, avoiding the wavelengths where the possible absorption lines from the background LBG and the foreground H~{\sc i} absorbers are contaminated. We estimated the flux uncertainty associated with each flux point by measuring the $1 \sigma$ of the background noise fluctuations at the same wavelength in the two-dimensional spectrum. The Voigt function fit for the observed gal-DLA spectrum was performed with three free parameters: redshift $z_{DLA}$, H~{\sc i} column density $N_{\rm HI}$, and continuum level $F_{\nu}^{cont}$. An example of the fit is shown in Figure~\ref{fig:voigtfit}. We repeated the fit by changing slightly the flux points used: increasing/reducing four or fewer contiguous flux points at the Ly$\alpha$ damped wing and including/excluding the eight flux points at $\lambda \sim 5150$\,\AA. We finally obtained the averaged best-fit parameters to be $z_{\rm DLA} = 3.335 \pm 0.007$ and $\log (N_{\rm HI}/{\rm cm}^{-2}) = 21.68 \pm 0.17$. The uncertainty for each quantity includes both the $1\sigma$ confidence interval ($\Delta \chi^2 < 1$) and the small scatter among the different settings of the flux points used. This column density is significantly larger than the threshold of DLAs ($\log(N_{\rm HI}/{\rm cm}^{-2}) = 20.3$). Our best-fit model spectrum also reproduces the possible Ly$\beta$ absorption line from the gal-DLA (Figure~\ref{fig:voigtfit}), ensuring the validity of our fitting. 

While the continuum of the background LBG is detected with high significance ($6 \sigma$ at $\lambda \approx 6000$\,\AA), no significant metal absorption line associated with the gal-DLA is identified. We put the $3 \sigma$ upper limit on the equivalent width (EW) of the gal-DLA metal absorption lines, resulting in EW$_0 < 1.4$\,\AA\ in the rest frame. This upper limit on the EW, which exceeds the observed range for QSO-DLAs \citep{Lu+96,Kaplan+10}, means only that the gal-DLA is not extremely metal-enriched. We show some possible DLA metal absorption lines in Figure~\ref{fig:DLAspec}, which should be confirmed with a deeper and higher resolution spectroscopy to further constrain the metal contents of the gal-DLA.

\subsection{Covering fraction of the H~{\sc i} gas cloud} \label{sec:DLAsize}

One of the merits in investigating gal-DLAs is that we may obtain constraints on the transverse extent of DLAs thanks to the spatial extension of the background light source. As it is impossible to spatially resolve the two-dimensional spectrum of the background LBG due to both of the S/N and seeing, we constrained the covering fraction of the gal-DLA over the background LBG by measuring the residual flux in the Ly$\alpha$ trough. As the observed residual flux at $\lambda = 5242 - 5295$\,\AA, which corresponds to the 10 pixels centered at the gal-DLA Ly$\alpha$ absorption, is consistent with zero, $F_{\nu,\rm res} = -0.009 \pm 0.05$\,$\mu$Jy, we put the $2 \sigma$ upper limit ($F^{\rm upp}_{\nu,\rm res} = 0.1$\,$\mu$Jy). The estimated lower limit of the covering fraction is $f^{\rm low}_{\rm cov} = 1 - F^{\rm upp}_{\nu,\rm res} / F_{\nu,\rm cont} = 0.7$, where $F_{\nu,\rm cont}$ is the continuum flux estimated in the section~\ref{sec:DLAgas}. 

By multiplying the lower limit of the covering fraction and the background LBG area, we can obtain the lower limit of the gal-DLA area projected to the background LBG plane. Since the background LBG is not resolved well in the ground-based images as shown in Figure~\ref{fig:stampimg}, we cannot measure the size directly. Thus, we used the bivariate size-luminosity relation in the rest UV frame for the $z = 3.4 - 4.4$ LBGs of \citet{Huang+13}. We assume an effective radius for the $z = 3.6$ background LBG from the median of the size distribution of the LBGs with the same UV luminosity, resulting in $1.6^{+2.1}_{-0.9}$\,kpc in physical scale. The upper and lower limit corresponds to the $16$ and $84$ percentiles, respectively (i.e., central $68$\,\% interval). We adopt the lower limit of the background LBG radius, $R_{\rm LBG} = 0.7$\,kpc, to estimate a conservative lower limit of the gal-DLA area. Assuming that the background LBG has the circular area, we estimated the lower limit of the DLA area by calculating $\pi R_{\rm LBG}^2 \times f^{\rm low}_{\rm cov}$. We finally convert the area in the background LBG plane to that in the gal-DLA plane, resulting in the lower limit of the gal-DLA area of $\sim 1$\,kpc$^2$. 

We compared our size estimate for the gal-DLA with dense H~{\sc i} gas clouds in galaxies produced in numerical simulations. \citet{Bird+13} examined the stacked radial H~{\sc i} density profile in the halos with $3\times 10^9\,h^{-1}\,M_{\sun} < M_{halo} < 3.5 \times 10^9\,h^{-1}\,M_{\sun}$ at $z = 3$ in the cosmological simulations (see their Figure~3), from which we can infer the DLA size as $R \approx 3 - 5\,h^{-1}$\,kpc in comoving scale or $R \approx 1.1 - 1.8$\,kpc in physical scale. For larger mass halos, cross sections of DLAs are clearly larger than $1$\,kpc$^2$ in physical scale \citep{Pontzen+08,Bird+13,Rahmati+14}. Therefore, our gal-DLA area ($\ga 1$\,kpc$^2$) is consistent with the size of DLAs expected from these simulations. 
 
We also compared our result with the DLA size measured using the gravitationally lensed QSO pairs. \citet{Cooke+10} analyzed the 20 gravitationally lensed QSO pairs and obtained the typical radius of DLAs at $z \sim 1.6$ as $R \approx 5 \pm 3$\,kpc in physical scale. Our estimate for the gal-DLA, $\ga 1$\,kpc$^2$, is compatible with the typical size of the DLAs along the lensed QSOs. The size estimation method in studies of lensed QSO pairs is sensitive to the maximum extents of DLA clouds because, generally, DLA absorptions can be seen in either of the pair sight-lines \citep{Monier+09,Cooke+10}. In contrast, our method constrains the minimum extent of DLAs because the area available in investigating the DLA size is absolutely limited by the extent of the background light source. The two methods are complementary, and combining their results yields more reliable estimates of DLA size.

\subsection{Counterpart galaxy of the gal-DLA}

\begin{figure}[]
%\begin{figure}[t]
\begin{center}
\includegraphics[width=1.0\linewidth, angle=0]{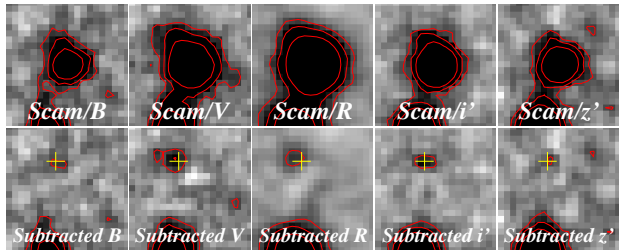}
\caption{Top five panels show multiband stamp images of the background LBG at $z = 3.604$, where the 2, 4, and $8 \sigma$ contours for each image are superposed. Each panel size is $4'' \times 4''$. North is up, and east is left. Bottom panels are the same as the top panels, except that the symmetrical component of the background LBG is subtracted from each image. The position of the marginal detection on the subtracted $i'$-band image is marked by the yellow cross. \label{fig:stampimg}}
\end{center}
\end{figure}

\begin{figure}[]
%\begin{figure}[t]
\begin{center}
\includegraphics[width=1.0\linewidth, angle=0]{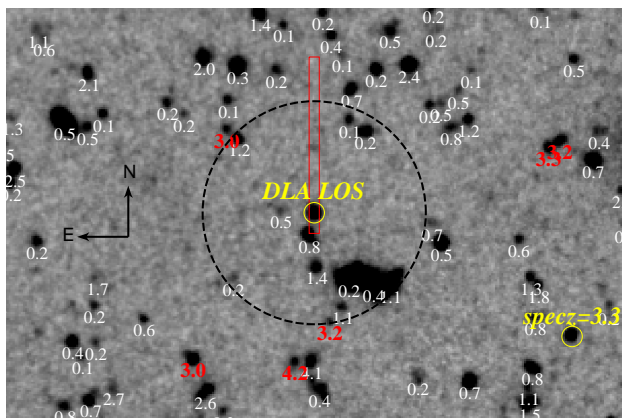}
\caption{The Suprime-Cam $i'$-band image centered at the position of the background LBG. The panel size is $20'' \times 30''$, which corresponds to physical 150\,kpc $\times$ 230\,kpc at $z = 3.3$. The red box shows the VIMOS slit for the background LBG. The background LBG and the closest confirmed LBG at $z = 3.303$ are shown by the yellow circles. Values below objects are the photometric redshifts, where those of the counterpart galaxy candidates (i.e., their $1\sigma$ redshift range include $z = 3.3$) are in bold. The dashed circle with the radius of $11''$ shows the area where no counterpart candidate at $z_{phot} \sim 3.3$ lies down to $i' = 26.4$. \label{fig:largeimg}}
\end{center}
\end{figure}

% counterpart
We searched for the Ly$\alpha$ line from the galaxy hosting the gal-DLA (hereafter the counterpart galaxy) in the Ly$\alpha$ trough in the background LBG spectrum (red box in Figure~\ref{fig:DLAspec}). This allows us to constrain the Ly$\alpha$ line flux of the counterpart galaxy uniformly inside the VIMOS slit ($1'' \times 17.4''$), where the slit configuration is shown in Figure~\ref{fig:largeimg}. We found no Ly$\alpha$ emitter (LAE) down to $F_{\rm Ly\alpha} = 1.9 \times 10^{-18}$\,erg\,s$^{-1}$\,cm$^{-2}$ ($3\sigma$), where the line flux is integrated over 4 spatial pixels and 5 spectral pixels ($\sim$ resolution size). The Ly$\alpha$ upper limit corresponds to the SFR of $\sim 2.5\,M_{\sun}$\,yr$^{-1}$, assuming the relation between the H$\alpha$ luminosity and SFR \citep{Kennicutt+98} with the correction for our adopted IMF \citep{Kennicutt+12}, the case B approximation \citep{Brocklehurst+71}, and Ly$\alpha$ escape fraction of $f_{\rm esc,Ly\alpha} = 0.05$ which is an average value for all galaxy populations at $z \sim 3$ \citep{Blanc+11,Hayes+11}. 

For objects fainter than the VIMOS detection limit or outside the VIMOS slit, we searched for the counterpart galaxy with the multiband imaging data: $u^*$ band from CFHT/Megacam \citep{Kousai+11}; $B, V, R_c, i', z'$ bands from Subaru/Scam \citep{Hayashino+04,Nakamura+11}; and $J and K$ bands from UKIRT/WFCAM (UKIDSS DXS DR10; \citealt{Casali+07,Lawrence+07}). The background LBG is marginally resolved in the Scam images and it seems to be elongated north-east (Figure~\ref{fig:stampimg}). We subtracted a smooth symmetric component from each Scam image to isolate any sub-components which may be the counterpart galaxy of the gal-DLA. The smooth symmetric component was made by stacking the objects with an FWHM similar to that of the background LBG in each image. There is a sub-component with $\sim 2 \sigma$ significance in the $i'$ and $V$ bands, and a less significant object at the same position in other bands (Figure~\ref{fig:stampimg}). While this object is a good candidate of the counterpart galaxy, its faintness makes it difficult for us to conclude whether this object is associated with the background LBG or the foreground gal-DLA. 

At larger distance, we searched for the counterpart galaxy using a standard photometric redshift technique. We constructed a photometric redshift catalog of all objects with $i' \le 26.6 (5\sigma)$. The photometric redshifts of objects around the gal-DLA are superposed on the $i'$-band image in Figure~\ref{fig:largeimg}. We consider the objects whose photometric redshifts are consistent with $z_{phot} = 3.3$ within the $1\sigma$ uncertainties as the counterpart galaxy candidates. The nearest candidate of the counterpart galaxy lies at $b = 11''$ (physical $84$\,kpc projected at $z = 3.3$) from the gal-DLA. This separation is considerably more distant compared to the previously reported counterpart galaxies ($\lesssim 25$\,physical kpc; \citealt{Krogager+12}), while the previous searches were biased to counterpart galaxies at smaller impact parameters as pointed out in \citet{Fumagalli+15}. Similarly, a spectroscopically confirmed LBG at $z = 3.303 \pm 0.008$ (Figure~\ref{fig:largeimg}) is not likely to be the counterpart galaxy because of the large redshift offset ($\Delta z = 0.030$, corresponding to physical 6.4 Mpc) and projected distance (physical 210\,kpc at $z = 3.3$). Assuming a fainter counterpart lying within $b = 11''$, we constrain the SFR of the counterpart galaxy to $< 0.8\,M_{\sun}$\,yr$^{-1}$, which is estimated from the $3\sigma$ of noise fluctuations in the $R_c$ band image with $2.2'' \phi$ $(2 \times {\rm FWHM})$ apertures and the relation between the UV luminosity and SFR \citep{Madau+98} with the correction for our adopted IMF \citep{Madau+14}. This constraint on the SFR is consistent with the previously reported SFRs of the galaxies hosting QSO-DLAs \citep{Fumagalli+15}, and then it is still possible that a modestly star-forming galaxy hosts the gal-DLA. We also searched for a UV-faint counterpart galaxy in the shallow $J$ and $K$ band images, and found no counterpart down to $J$ or $K \sim 23$\,mag, which means that neither a passive galaxy with $M_* \gtrsim 5 \times 10^{10}\,M_{\sun}$ nor a dusty star-forming galaxy with $E(B - V) \gtrsim 2$ lies near the gal-DLA.

%%%%%% long version %%%%%%%%%%%
\subsection{Environment around the gal-DLA}

\begin{figure}[]
%\begin{figure}[t]
\begin{center}
\includegraphics[width=1.0\linewidth, angle=0]{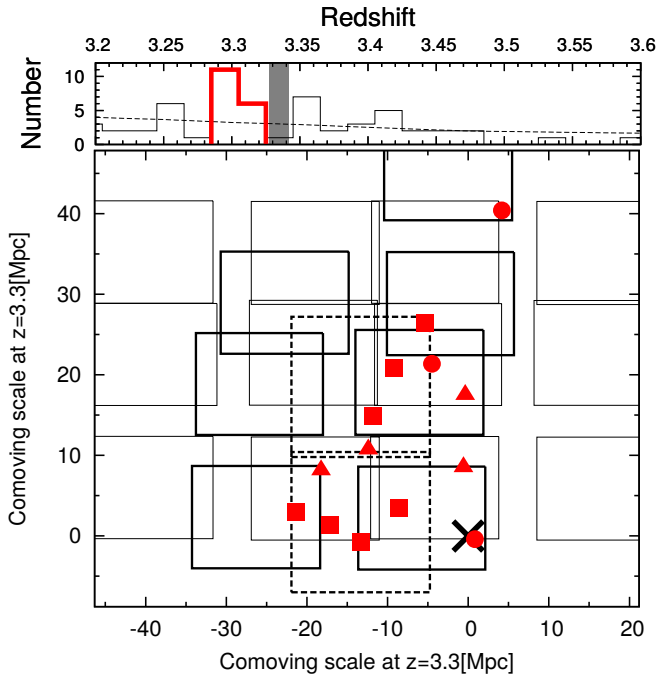}
\caption{Top panel shows the redshift distribution of the LBGs, which are gathered from our observation, \citet{Steidel+03}, and \citet{Kousai+11}. The expected number histogram (T. Hayashino et al. 2016, in preparation) is shown by the dotted curve. The redshift of the gal-DLA is shown by the gray shading where the width corresponds to the $1 \sigma$ uncertainty. The possible density peak at $z = 3.28 - 3.32$ is marked by the red thick histogram. Bottom panel shows the sky distribution of the gal-DLA (black cross) and the LBGs at $z = 3.28 - 3.32$ (circles: our observation, squares: \citealt{Steidel+03}, and triangles: \citealt{Kousai+11}). The FOVs of our observation, \citet{Steidel+03}, and \citet{Kousai+11} are shown by the thick solid, thick dashed, and thin solid line, respectively. \label{fig:skymap}}
\end{center}
\end{figure}

% DLA environment
It have been suggested that the galaxy overdensity environment is responsible at least for some QSO-DLAs \citep{Bouche+03,Chen+03,Cooke+06b,Kacprzak+10}. We examine the galaxy density environment around the gal-DLA in this section. From our spectroscopic LBG sample we found a possible overdensity at $z \sim 3.3$, which can also be seen in the observations by \citet{Steidel+03} and our previous VIMOS survey (VIMOS06; \citealt{Kousai+11}). The redshift distribution of the LBGs observed in the three surveys (VIMOS08, VIMOS06, and \citealt{Steidel+03}) and the sky distribution of the LBGs at $z = 3.28 - 3.32$ are shown in Figure~\ref{fig:skymap}. The number overdensity of the LBGs at $z = 3.28 - 3.32$ ($\delta_{\rm LBG}$) is $\approx 1.6 \pm 0.6$, where the number expected from the LBG selection functions of the observations was used as the average. In the sky distribution, the LBGs at $z = 3.28 - 3.32$ seem to be assembled in a portion of the whole observation coverage, although we should take into account the difference in the FOVs and depth of the surveys. Then, the number overdensity of the LBG structure may be larger than $\delta_{\rm LBG} \approx 1.6$, which is the estimate for the whole observation area. We evaluated the finding probability of this LBG large-scale structure in the $z = 3.3$ universe, following the method introduced in \citet{Mawatari+12} with the $z \sim 3$ LBG linear bias of $b_{\rm LBG} = 3$ \citep{Lee+06}. We adopted $\delta_{\rm LBG} = 1.6$ conservatively and assumed that the LBG overdensity region is a spheroid with $r = 20$\,Mpc (comoving). The resulting finding probability is $1.6^{+5.8}_{-1.3}$\,\%, suggesting that the LBG overdensity region is a relatively rare galaxy group in the $z = 3.3$ universe. On the other hand, the redshift of the gal-DLA, $z = 3.335 \pm 0.007$, is slightly offset from the possible density peak in the redshift distribution. There are some artificial effects like the physical constraints of the slit configuration and redshift uncertainties, and we cannot conclude whether the gal-DLA is related to the possible LBG overdensity. A more complete observation such as an integral field unit spectroscopy or a narrow-band imaging survey is needed to investigate the environmental dependency of the gal-DLA.

%%%%%%%%%%%%%%%%%%%%%%%%

\subsection{The gal-DLA finding probability}

We found one gal-DLA among the 80 LBG spectra, resulting in a frequency of $\approx 1$\,\% for DLAs with $\log(N_{\rm HI}/cm^{-2}) \ga 21.68$. We compare this observed frequency with that expected from the published DLA number density. We here assume that gal-DLAs exhibit an identical incidence to QSO-DLAs, which is not obvious. If the size of a DLA projected onto a background galaxy is much smaller than the size of the galaxy, there should be a significant residual flux of the background galaxy at the bottom of the line profile (see Sec~\ref{sec:DLAsize}), and then the absorption may not be recognized as a DLA along the galaxy sight-line. In this case, the DLA occurrence rate along galaxy sight-lines would be smaller than those along a compact source like QSOs and GRBs.

We used the distribution function (i.e., the number density per unit H~{\sc i} column density per unit redshift) of the intergalactic absorbers of \citet{Inoue+14} to calculate the expected number of DLAs in the 80 LBG spectra: 
\begin{eqnarray*}
n_{DLA} = \sum^{80}_{i=1} \int^{z_{max,i}}_{z_{min,i}}\int^{\infty}_{N_{min}} \frac{\partial^2 n}{\partial z \partial {N_{\rm HI}}} dz dN_{\rm HI}~,
\end{eqnarray*}
where $i$ is the index of each LBG at $z = z_i$. The minimum H~{\sc i} column density is set to the $2 \sigma$ lower limit of the observed value, $N_{\rm min} = 2.2 \times 10^{21}$\,cm$^{-2}$. The maximum and minimum redshift used for the gal-DLA search in each LBG spectrum are set as $z_{{\rm max},i} = 1170(1 + z_i)/1216-1$ and $z_{{\rm min},i} = 1070(1 + z_i)/1216-1$ because we inspected the wavelength range corresponding to $\lambda = 1070 - 1170$\,\AA\ in each LBG rest frame. We selected this range to isolate the intergalactic H~{\sc i} Ly$\alpha$ absorption avoiding the S~{\sc iv}~$\lambda$1063 and C~{\sc iii}~$\lambda$1178 absorption lines from the LBGs themselves. For three among the 80 LBGs we instead adopted $z_{{\rm min},i} = 3800/1216-1$, where the observed wavelengths corresponding to rest-frame 1070\,\AA\ in their spectra are shorter than the short edge of the VIMOS observable wavelength range ($\sim 3800$\,\AA). We finally obtained the expected number of gal-DLAs with $\log(N_{\rm HI}/cm^{-2}) > 21.34$ in our survey of $0.26$ or an expected frequency of $\approx 0.33$\,\%. 

A finding probability of one gal-DLA with $\log(N_{\rm HI}/{\rm cm}^{-2}) \geq 21.34$ ($2 \sigma$ lower limit) in our survey is $\approx 20$\,\% assuming a Poisson probability distribution with the expected value of $0.26$. This modest probability suggests that our finding of 1 gal-DLA among the 80 spectra is a relatively lucky event but it can be explained within the occurrence rate for QSO-DLAs. In other words, our assumption that the occurrence rate for DLAs does not depend on their background sources is not rejected. 

We can expect that the number of gal-DLAs with the similar to or lower H~{\sc i} column density than the observed gal-DLA in this study will increase with archival and future spectroscopic observations, owing to the steep slope of the DLA distribution function with respect to $N_{\rm HI}$ \citep{Prochaska+05,Noterdaeme+09,Noterdaeme+12b}. The statistical study of gal-DLAs will open a new window on examining the DLA size and properties of the counterpart galaxies such as the SFR and the stellar mass. With a large statistical sample of gal-DLAs, we can discuss the differences in the DLA occurrence rates depending on the background source, yielding a further constraint on the DLA size.


%% Included in this acknowledgments section are examples of the
%% AASTeX hypertext markup commands. Use \url without the optional [HREF]
%% argument when you want to print the url directly in the text. Otherwise,
%% use either \url or \anchor, with the HREF as the first argument and the
%% text to be printed in the second.

\acknowledgments

This work is based on observations collected at the European Organisation for Astronomical Research in the Southern Hemisphere under ESO programme (081.A-0081) and at the Subaru Telescope which is operated by the National Astronomical Observatory of Japan. We would appreciate Jeff Cooke and John O'Meara showing their findings of the first gal-DLA example and discussing it with us about it prior to its publication. K.M. and A.K.I. also appreciate Nobunari Kashikawa and Toru Misawa for their helpful comments. This work was financially supported by JSPS KAKENHI Grant Number 26287034.


%% in parentheses, after the keyword, but they will not be verified.

{\it Facilities:} \facility{VLT (ESO)}, \facility{Subaru (NAOJ)}.

\end{document}